# Lattice dynamics of MC$_{60}$ compounds in FCC phase

K. Ranjan[1], N. Kaurav[2], Dinesh Varshney[2], Keya Dharamvir[1] and V. K. Jindal[1,*]

[1] Department of Physics, Panjab University, Chandigarh, India-160014

[2] School of Physics, Devi Ahilya University, Khandwa road Campus, Indore, India – 452001

## Abstract

*Phonon dynamics of alkali metal (M) doped C$_{60}$ ( MC$_{60}$ ) solids in FCC phase has been studied. The calculations take into account van der Waal's and Coulomb interactions for alkali metal− C$_{60}$ as well C$_{60}$–C$_{60}$ molecule and show fairly good agreement with reported neutron scattering results for RbC$_{60}$. The calculations have also been done following Rigid Shell Model (RSM). We also perform the calculations for specific heat, Gruneisen parameter, thermal expansion and thermal expansion coefficient.*

Keywords: Dynamics; Phonons; Ionic solid; Thermodynamics; Specific heat; MC$_{60}$; Fullerenes

PACS:  61.48.+c, 62.20.Dc, 63.20.Dj, 65.20.+w

## 1. Introduction

Among alkali doped C$_{60}$ solids the composition MC$_{60}$ (M = Na, K, Rb and Cs) exhibit a variety of stable phases and are of considerable interest in recent past. X-ray diffraction studies have shown that at high temperatures these compounds exist in a rock salt structure [1] which transforms to an orthorhombic phase [2,3] upon moderate cooling to room temperature. On cooling to liquid nitrogen temperature they transform to a monoclinic phase [4]. Other phases have been found upon warming from the lower temperature phases [5]. In the present work we have calculated phonon dispersion curves in the symmetry directions for FCC phase. The calculation is based upon our earlier work [6-8] where we have established a model to account for bulk properties of alkali doped C$_{60}$ solids assuming C$_{60}$ as a continuous shell of radius 3.55 Å. The FCC rock salt structure of MC$_{60}$ is formed with freely rotating C$_{60}^-$ ions above 370 K [5]. Therefore it would be justifiable to calculate phonon dispersion curves and other related quantities of the FCC phase of MC$_{60}$ using the continuous model [6-8]. It has been established that these compounds are ionic [6] in character. In the present work we perform the analytical calculations for external phonons by considering van der waal's attractive and Born Mayer repulsive potential upto second nearest neighbor, where as for the

---

[*] Corresponding Author: E-mail: jindal@pu.ac.in

Coulomb part of the dynamical matrix we have used the quantities evaluated by Kellermann [9] using the Ewald sum method. We have done calculations in the frame work of Rgid Ion Model (RIM) and Rigid Shell Model (RSM) at equilibrium lattice constant determined by us [7,8] and with slight modification of these interaction parameters so as to match lattice constant value with experimental value for all M (Na, K, Rb and Cs). The density of states in case of $RbC_{60}$ has been compared with available Neutron scattering results [10]. We have also calculated elastic constants, long wave optic phonons, specific heat, thermal expansion and thermal expansion coefficient for the whole series of these compounds. In section 2 we discuss the models used. Results and discussion are presented in section 3.

## 2. Theoretical Formalism

### 2.1 The model for lattice dynamics

In case of usual ionic solids, nearest neighbor overlap potential (repulsive) and Coulomb part is enough to describe the structure. In the present system however the anion is a large cluster of 60 carbon atoms. Therefore not only repulsive but attractive van der Waal's term is also important to find force constants. We have earlier worked out [6] the form of potential between two $C_{60}$ molecules assuming that the potential between two carbon atoms is given by a 6-exp potential,

$$V_{6-\exp} = -\frac{A}{r^6} + B e^{-\alpha r}. \tag{1}$$

Further $C_{60}$ molecule is modeled as a spherical shell of radius RB (the "continuum approximation") on which 60 carbon atoms are uniformly smeared over the surface. This allows us to express potential between two $C_{60}$ molecules in term of carbon-carbon potential parameters. The form of the $C_{60}$-$C_{60}$ potential is

$$\Phi_{BB}^{6-\exp}(R) = -(60)^2 \left[ \left(\frac{A}{R^6}\right) \frac{1 - 2(R_B/R)^2 + 8/3(R_B/R)^4}{\{1-(2R_B/R)^2\}^3} - Be^{-\alpha R}\left(\frac{\sinh(\alpha R_B)}{\alpha R_B}\right)^2 \left\{1 + \frac{2}{\alpha R}\left(1 - \frac{\alpha R_B}{\tanh(\alpha R_B)}\right)\right\} \right] \tag{2}$$

where R is the distance between the centers of two $C_{60}$ molecules. Assuming similar atom-atom potential ( Eqn. 1) for alkali metal and carbon; the potential between alkali metal and $C_{60}$ is given by

$$\Phi_{BM}^{6-\exp}(r) = -\frac{60 A}{r^6}\left[\frac{\left(1+\left(\frac{R_B}{r}\right)^2\right)}{\left(1-\left(\frac{R_B}{r}\right)^2\right)^4}\right] + \frac{60 \cdot B \cdot e^{-\alpha r}}{R_B \alpha}\left[-\frac{R_B}{r}\cosh(R_B \alpha)+\left(1+\frac{1}{r\alpha}\right)\sinh(R_B \alpha)\right] \tag{3}$$

where $r$ is the centre to centre distance between alkali metal ion and $C_{60}$ molecule. $A$, $B$ and $\alpha$ are the 6-exp potential parameters for alkali metal and carbon atom.

The cohesive energy per unit cell is given by

$$U_0 = -\alpha_M Z^2 \frac{e^2}{r_0} + 6\Phi_{BM}^{6-\exp} + 12\Phi_{BB}^{6-\exp} \qquad (4)$$

Here, $\alpha_M$ is the Madelung constant for $MC_{60}$ which is in NaCl structure. $2r_0$ is the lattice constant of the (cubic) unit cell and $Z$ the number of electronic charge on the positive ion. In Eqn. 4, the multiplier 6 and 12 in second and third term are number of nearest neighbors and second nearest neighbors respectively. Two parameters have been defined in terms of the derivatives of the potential defined in Eq. (2) and Eq. (3) as

$$B_1 = \frac{2v}{e^2}\left[\frac{1}{R}\frac{d}{dR}\Phi(R)\right]_{R=R_0} \qquad (5a)$$

and

$$A_1 = \frac{2v}{e^2}\left[\frac{d^2}{dR^2}\Phi(R)\right]_{R=R_0} \qquad (5b)$$

where $v = 2r_0^3$ is the volume of the primitive cell and $R_0$ is the equilibrium separation between the ions for which $B_1$ and $A_1$ is defined. In view of this, $R_0 = r_0$ for alkali metal and $C_{60}$ and $R_0 = r_0 \times \sqrt{2}$ for $C_{60}$ and $C_{60}$. The potential $\Phi(R)$ is the potential defined in Eq. (2) and Eq. (3) as the case may be depending upon the type of ions taken. Therefore we have set of parameters $A_1$ and $B_1$ for similar ions (second neighbor) and for dissimilar ions (first neighbor). While considering second neighbor the contribution of M-M has been neglected as this interaction potential is almost $\frac{1}{60} \times \frac{1}{60}$ times compared to $C_{60}$-$C_{60}$ interaction potential (please see Eqn. 2).

According to the picture of Rigid Ion Model (RIM) [11], the potential energy of interaction of a pair of ions separated by a distance $r$ is given by

$$\Phi_{kk'}(r) = \Phi_{kk'}^C(r) + \Phi_{kk'}^{6-\exp}(r) = Z_k Z_{k'} e^2 {1}/{r} + \Phi_{kk'}^{6-\exp}(r) \qquad (6)$$

where $Z_k e$ is the charge of the ion in the $k^{th}$ sublattice and $e$ is the magnitude of electronic charge. This potential correspond to a dynamical matrix [9] of the following form

$$D_{\alpha\beta}(kk'/\vec{q}) = D_{\alpha\beta}^C(kk'/\vec{q}) + D_{\alpha\beta}^{6-\exp}(kk'/\vec{q}) = Z_k e C_{\alpha\beta}(kk'/\vec{q}) + R_{\alpha\beta}(kk'/\vec{q}) \qquad (7)$$

where α and β denote the Cartesian component indices; $\vec{q}$ is the wave vector associated with the wave. It may be expressed compactly in matrix notation as

$$D(\vec{q}) = ZCZ + R \tag{8}$$

where Z is a ($3n \times 3n$) diagonal matrix with elements

$$Z_{\alpha\beta}(kk') = Z_k e \delta_{\alpha\beta} \delta_{kk'} \tag{9}$$

and C and R are ($3n \times 3n$) diagonal matrices with elements

$$C_{\alpha\beta}(kk'/\vec{q}) = \sum_{l'} \Phi^C_{\alpha\beta}(lk; l'k') \exp[i\vec{q}\cdot\vec{r}(lk; l'k')] \tag{10}$$

$$R_{\alpha\beta}(kk'/\vec{q}) = \sum_{l'} \Phi^{6-\exp}_{\alpha\beta}(lk; l'k') \exp[i\vec{q}\cdot\vec{r}(lk; l'k')] \tag{11}$$

where $\vec{r}(lk, l'k')$ is the distance between two atoms whose position is represented by ($lk$) i.e. $l$ th unit cell and $k$ th atom and $(l'k')$ i.e. $l'$ th unit cell and $k'$ th atom. The general definition of force constant $\Phi_{\alpha\beta}$ defined in term of ion - ion pair potential is given in reference 12. The matrix elements in Eq. 11 contain the direct lattice sums which converge very fast due to van der Waals and Born Mayer interaction. In fact in binary ionic solids (NaCl, KBr, NaI etc.), it is only the repulsive part which is important, therefore it is justified and sufficient enough to take first neighbour interaction to evaluate Eq. (11). However in case of molecular solid (like $MC_{60}$) to evaluate $R_{\alpha\beta}$ at least one should take contribution upto second neighbors, so that $C_{60}$-$C_{60}$ repulsion, which is dominant, is taken into account. The attractive part may not be that important as the solid is ionic. Keeping in view the size of $C_{60}$ we have retained attractive part and hence the name "$\Phi^{6-\exp}_{\alpha\beta}$".

In case of Eq. 10 the sum is over Coulomb potential. For Coulomb potential series is not convergent and hence direct lattice sum is not possible. It is worth mentioning here that Evjen's Method [6] of direct lattice sum is useful to calculate Madelung energy, but it does not work for the directional summation involved in Eq. 10. This is why we have used the matrix elements which are calculated by Kellermann [9] using Ewald sum technique. The R matrix elements may be written for rock salt structure in terms of force constants (defined by Eqns. 5a and 5b) taking the contribution upto second neighbor [11] as

$$R_{\alpha\alpha}(kk') = -\frac{e^2}{v}[A_1' Cos\,\pi q_\alpha + B_1'(Cos\,\pi q_\beta + Cos\,\pi q_\gamma)] \tag{12a}$$

$$R_{\alpha\alpha}(kk) = R_0 + \frac{e^2}{v}[(A_1'' + 2B_1'')$$
$$-\frac{A_1'' + B_1''}{2} Cos\,\pi q_\alpha (Cos\,\pi q_\beta + Cos\,\pi q_\gamma) - B_1'' Cos\,\pi q_\beta Cos\,\pi q_\gamma]$$
(12b)

$$R_{\alpha\beta}(kk) = \frac{e^2}{v}[(A_1'' - B_1'')\,Sin\,\pi q_\alpha Sin\,\pi q_\beta)];\ R_{\alpha\beta}(kk') = 0 \tag{12c}$$

$$R_0 = \frac{e^2}{v}(A_1' + 2B_1') \tag{12d}$$

where the single primed quantities ($A_1'$ and $B_1'$) correspond to the Metal and $C_{60}$ (first neighbour) interaction, whereas double primed quantities ($A_1''$ and $B_1''$) correspond to either $C_{60}$-$C_{60}$ (second neighbor) or M-M interaction. The elements thus calculated are added to the *C* matrix elements and hence the total dynamical matrix is evaluated for 48 points in the Brillioun zone. Diagonalization of the same lead to frequencies as a function of $\vec{q}$, the wave vector and the total number of frequencies thus calculated is $3sN_q$. $N_q$ is the total no. of $\vec{q}$ vectors and s the number atoms in the primitive cell.

We have also performed calculations with Rigid Shell model (RSM). In brief RSM [14] postulates the ion to be composed of a central massive core consisting of nucleus and tightly bound electrons, attached by harmonic springs to an outer massless spherical shell made of valence electrons. The shell retains its spherical charge but can move bodily with respect to its core. This relative mechanism gives rise to electronic polarization. The valence electron cloud (shell) of anion increases compared to the neutral atom whereas in the case of cation the shell either decreases in size or just ceases and consequently only core is left (e.g. in the case of alkali metals). Therefore the polarizability of the negative ion (anion $C_{60}$) is more important compared with positive ion (alkali metal). With single ion polarizable, the complex treatment in RSM gets simplified and one can easily construct the dynamical matrix [13] to calculate the phonon frequencies.

## 2.2. The long wave approximation

We now calculate the elastic constants of the solid (or matrix elements $C_{ij}$ of elastic stiffness constant) in the continuum model of the solid.

### 2.2.1 The acoustic behavior

The expression for the elastic constants can be obtained with the help of standard procedure outlined in ref. 11. The expressions for various elastic constants thus obtained for rock salt structure are,

$$C_{11} = \frac{e^2}{4r_0^4}\left[-5.112Z^2 + A_1' + \frac{1}{2}\left(A_1'' + B_1''\right)\right] \quad (13)$$

$$C_{12} = \frac{e^2}{4r_0^4}\left[0.226Z^2 - B_1' + \frac{1}{4}\left(A_1'' - 5B_1''\right)\right] \quad (14)$$

$$C_{44} = \frac{e^2}{4r_0^4}\left[2.556Z^2 + B_1' + \frac{1}{4}\left(A_1'' + 3B_1''\right)\right] \quad (15)$$

In view of the equilibrium condition the RIM/RSM framework leads to the Cauchy relation ($C_{12} = C_{44}$). The polarizability has no effect on the elastic constants. We shall now proceed to seek the optical behaviour of the present molecular solid.

### 2.2.2 The optical behavior

The long-wave optical vibration frequencies can be derived with nearest neighbor approximation in the frame work of RIM and are given by

$$\omega^2 = \frac{R_0}{\mu}\left[1 + C_1 \alpha_I / v\right] \quad (16)$$

In RSM (negative ion polarizable), the frequencies are given [13] by

$$\mu\omega^2 = R_0 - \frac{e^2 d^2}{\alpha} + \frac{e^2}{v}\left[\frac{C_1(Z-d)^2}{1 + \alpha C_1/v}\right] \quad (17)$$

Here the symbol μ is the reduced mass and is $\mu = m_1 m_2 / (m_1 + m_2)$ and $\alpha_I = (Ze)^2 / R_0$ denotes ionic polarizability. The frequency of the transverse optic mode, $\omega_T$, may be obtained by setting $C_I = -4\pi/3$ and that of the longitudinal mode, $\omega_L$ by setting $C_I = 8\pi/3$. The parameter $d$ is defined [13] in terms of electrons in the shell of negative ion. Henceforth it is convenient to consider it as a parameter equivalent to Y, the electrons in the shell. We shall use Eq's. 16 and 17 to deduce the optical frequencies at zone centre.

### 2.3. Determination of the specific heat

Usually, the distribution of phonon frequencies are employed to compute the heat capacity at constant volume, $C_v$ as

$$C_v = \left[\frac{\partial E}{\partial T}\right]_V \quad (18)$$

where, internal energy E is given by

$$E = F - T\left[\frac{\partial F}{\partial T}\right]_V = \sum_{qj}\left\{\Phi(V) + \frac{\hbar\omega_{qj}}{2} + \frac{\hbar\omega_{qj}}{\exp[\hbar\omega_{qj}/kT] - 1}\right\} \quad (19)$$

Here, $F$ is the Helmholtz free energy function. Thus,

$$C_V = k\sum_{qj}\left[\left(\frac{\hbar\omega_{qj}}{2kT}\right)^2 \Big/ \sinh^2\frac{\hbar\omega_{qj}}{2kT}\right]\Big/N_q \quad (20)$$

## 2.4 Determination of thermal expansion coefficient

Minimizing the quasi-harmonic free energy of the crystal, the lowest order volume thermal expansion or dialation is given by [14]

$$\varepsilon = \frac{\Delta V}{V} = \frac{1}{2V_0 BM}\frac{1}{N_q}\sum_{qj}\gamma_{qj}\hbar\omega_{qj}\coth\frac{\hbar\omega_{qj}}{2kT} \quad (21)$$

where $V_0$ is the volume at zero temperature, $B$ is bulk modulus and $\gamma_{qj}$ is Gruneisen parameter defined as

$$\gamma_{qj} = -\left(\frac{\partial\ln\omega_{qj}}{\partial\ln V}\right) = -\frac{V}{\omega_{qj}}\frac{\partial\omega_{qj}}{\partial V} \quad (22)$$

The volume expansion coefficient $\alpha_{TH}$ is the derivative of dialation (Eq. 21) with respective to temperature and directly proportional to specific heat. The expression for the same is given as

$$\alpha_{TH} = \frac{\partial\varepsilon}{\partial T} = \frac{k}{V_0 B}\frac{1}{N_q}\sum_{qj}\gamma_{qj}\left(\frac{\hbar\omega_{qj}}{kT}\right)^2\frac{e^{\hbar\omega_{qj}/kT}}{\left(e^{\hbar\omega_{qj}/kT} - 1\right)^2} \quad (23)$$

## 2.5 The method of calculations

As mentioned earlier in section 2.1, we have earlier determined [6-8] parameters A, B and α for the interactions given by Eq.2 and Eq.3. The values of these parameters (say Set-1) are listed below in the Table 1 for the sake of completeness.

**Table 1: Interaction parameters [Set-1]**

| Atom-Atom | A (kCal-Å$^6$/mol) | B (KCal/mol) | α (Å$^{-1}$) |
|---|---|---|---|
| C-C | 358 | 42000 | 3.58 |
| Na-Na | 100 | 23730 | 4.45 |
| Na-C | 160 | 35000 | 3.60 |
| K-K | 171 | 49138 | 3.62 |
| K-C | 235 | 28370 | 3.50 |
| Rb-Rb | 333 | 76628 | 3.43 |
| Rb-C | 458 | 31283 | 3.32 |
| Cs-Cs | 495 | 114942 | 3.04 |
| Cs-C | 680 | 38314 | 3.12 |

The lattice constant values obtained with above parameters are slightly small compared with experimental results [17]. So we have taken another set of phonon dispersion curves with adjusting parameter B and α for Alkali metal-Carbon interaction. Other parameters remain unaltered. It is evident that this interaction is instrumental in the determination of lattice constant. The dispersion curves thus obtained have a different shape and the effect is more prominent in the optical branch. We have also done RSM calculations using these parameters. Table 2 gives the set of these parameters (Set-2).

**Table 2: Interaction parameters [Set-2]**

| Atom-Atom | A (kCal-Å$^6$/mol) | B (KCal/mol) | α (Å$^{-1}$) |
|---|---|---|---|
| C-C | 358 | 42000 | 3.58 |
| Na-Na | 100 | 23730 | 4.45 |
| Na-C | 160 | 60000 | 3.30 |
| K-K | 171 | 49138 | 3.62 |
| K-C | 235 | 70000 | 3.27 |
| Rb-Rb | 333 | 76628 | 3.43 |
| Rb-C | 458 | 80000 | 3.24 |
| Cs-Cs | 495 | 114942 | 3.04 |
| Cs-C | 680 | 100000 | 3.22 |

The values of lattice constant thus obtained using the calculation procedure explained in [6] are given in Table 3.

**Table No. 3: Lattice constant along with the reported data.**

| | Lattice constant (Å) Present Calculation | Lattice constant (Å) Reported data |
|---|---|---|
| NaC$_{60}$ | 14.04 | 14.14 [15]* |
| KC$_{60}$ | 14.07 | 14.07 [156] |
| RbC$_{60}$ | 14.08 | 14.08 [16] |
| CsC$_{60}$ | 14.12 | 14.12 [16] |

∗ Calculated value.

Using these two sets of parameters mentioned above we determine phonon frequencies in the frame work of RIM. The calculation in the frame work of RSM has been done only for Set-2. The mass of C$_{60}$ is taken to be 720 *a.m.u. i.e.* 1195×10$^{-24}$ g. Various other input parameters used in the calculations for Set-1 and for Set-2, are listed below in Table 4 and 5, respectively.

**Table 4 The input parameters for Set-1 (RIM)**

| | Nearest neighbor distance (10$^{-8}$cm) | $A_1'$ (CGS) | $B_1'$ (CGS) | $A_1''$ (CGS) | $B_1''$ (CGS) | Mass of alkali metal atom (10$^{-24}$g) |
|---|---|---|---|---|---|---|
| NaC$_{60}$ | 6.975 | 6.484 | -0.123 | 109.208 | -0.609 | 38 |
| KC$_{60}$ | 6.975 | 6.122 | -0.053 | 109.208 | -0.609 | 65 |
| RbC$_{60}$ | 6.975 | 11.824 | -0.147 | 109.208 | -0.609 | 142 |
| CsC$_{60}$ | 7.00 | 31.45 | -0.985 | 94.460 | -0.247 | 221 |

**Table 5 The input parameters for Set-2 (RIM)**

| | Nearest neighbor distance (10$^{-8}$cm) | $A_1'$ (CGS) | $B_1'$ (CGS) | $A_1''$ (CGS) | $B_1''$ (CGS) | Mass of alkali metal atom (10$^{-24}$g) |
|---|---|---|---|---|---|---|
| NaC$_{60}$ | 7.020 | 34.046 | -1.406 | 83.813 | 0.414 | 38 |
| KC$_{60}$ | 7.035 | 41.374 | -1.694 | 76.573 | 0.178 | 65 |
| RbC$_{60}$ | 7.040 | 48.530 | -1.869 | 74.238 | 0.232 | 142 |
| CsC$_{60}$ | 7.060 | 59.913 | -2.242 | 65.497 | 0.431 | 221 |

For RSM (negative ion polarizable) the value of parameter *d* is taken to be 0.1. The electronic polarizability of the negative ion i.e. C$_{60}^-$ α, is 84 ×10$^{-24}$cm$^3$ [17]. We have calculated frequencies with two sets of parameters (Set-1 and Set-2) as mentioned above at two different values of volume (at equilibrium and slightly deviated from equilibrium) so as to evaluate the thermodynamic quantities defined by Eq's. 21, 22 and 23, respectively.

## 3. Results and Discussion

In the whole series of $MC_{60}$ compounds, we have been able to compare our calculations with experiment only for $RbC_{60}$. Fig.1 shows the variation of calculated phonon density of states (DOS) for $RbC_{60}$ with frequency $\omega$ along with inelastic neutron scattering results [10]. To compare our calculations with the experiment, the DOS of the reference 10 has been normalized with the present calculations. In Fig.1, the cut-off frequency of $\approx$ 7.2 meV for set-1 is quite close to the experimentally observed $\approx$ 8 meV [10]. Earlier investigations [18] of $Rb_3C_{60}$ have also concluded that vibrational frequencies of octahedral Rb sites lie around 7 meV, which is in good agreement with present calculation results.

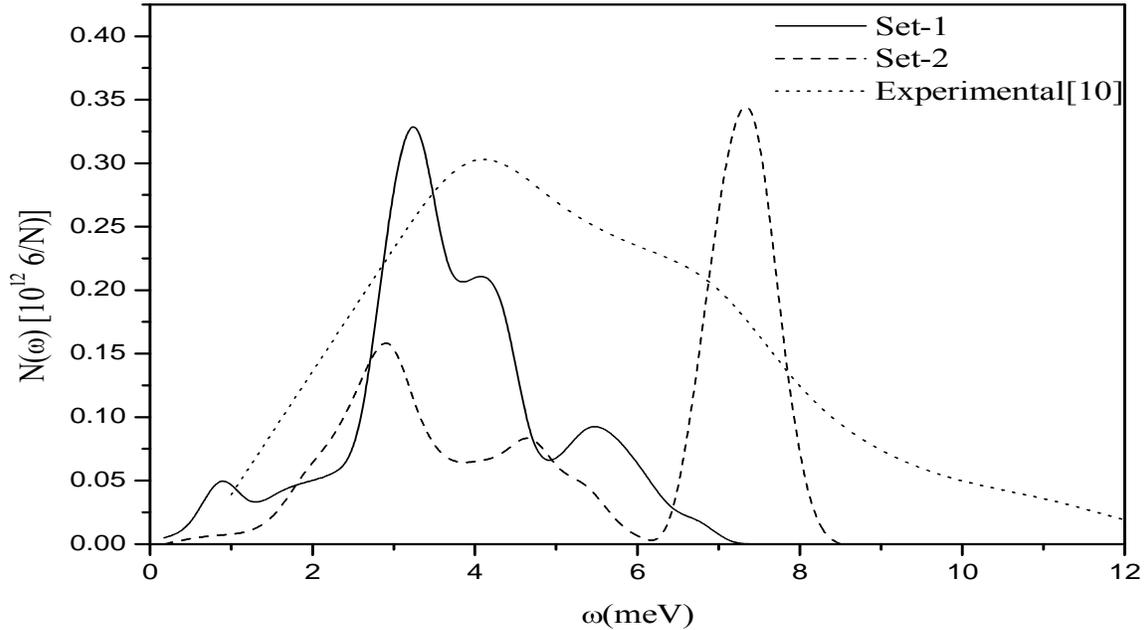

Fig.1. Comparision of calculated density of states (RIM) with the Experiment.

Further, the shape of calculated phonon DOS curve resembles the experimental curve. The DOS curve corresponding to Set-1 shows a maximum at 3.25 meV and a shoulder at 4.3 meV. However, compared with experimental DOS curve the maximum and shoulder in the present calculations are shifted towards the lower energy end. The DOS curve for Set-2 does not resemble with the experimental curve. Therefore the calculations with set-1 are more realistic and should predict thermodynamic quantities more correctly. It is important to mention that with set-2 the optical modes are well separated from acoustic modes (Fig.2 and Fig.3), which changes the characteristic of the spectrum. However, the cut-off frequency remains ≈ 8.5 meV, which is not far from experimental values mentioned above.

We have also done calculations in the frame work of Rigid Shell Model (RSM). Fig. 4 shows frequencies calculated along symmetry directions with set-2. Comparing with Fig.3 one readily finds that the phonon dispersion curves (PDC's) do not change significantly to affect the related thermodynamic quantities. Therefore further calculations are not done in this frame work.

Now it is established that the parameters corresponding to set-1, shall be the appropriate for dynamics of these compounds. It is worth mentioning here that these parameters of set-1 have already been used to determine various bulk properties of $Rb_nC_{60}$ solids [7].

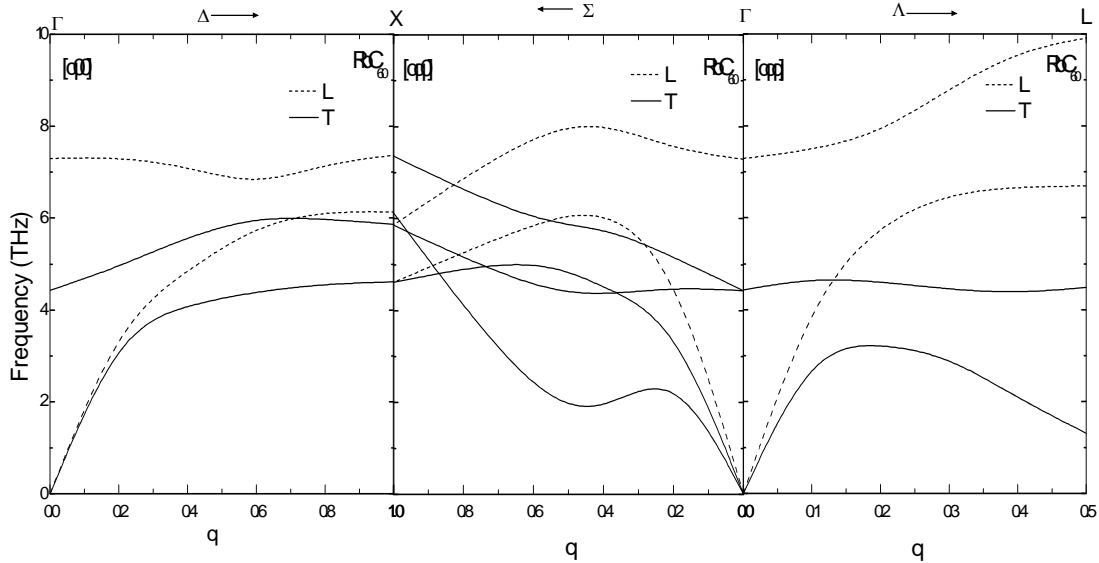

Fig. 2. Phonon dispersion curves (RIM) of $RbC_{60}$ in symmetry direction with Set-1

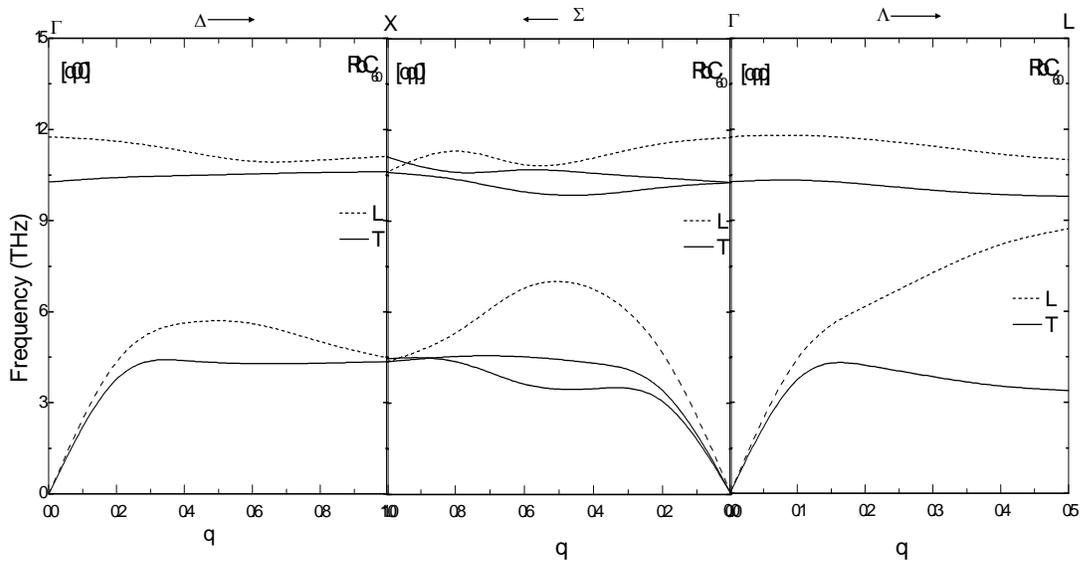

Fig. 3 Phonon dispersion curves (RIM) of RbC$_{60}$ in symmetry direction with Set-2.

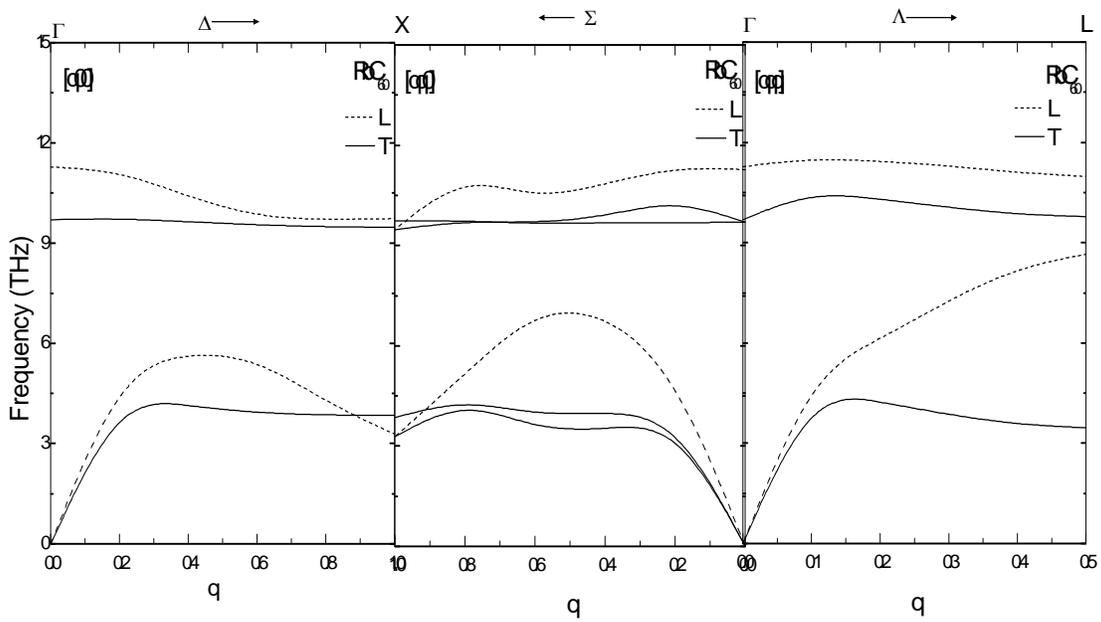

Fig.4 Phonon dispersion curves (RSM) of RbC$_{60}$ in symmetry direction with Set-2.

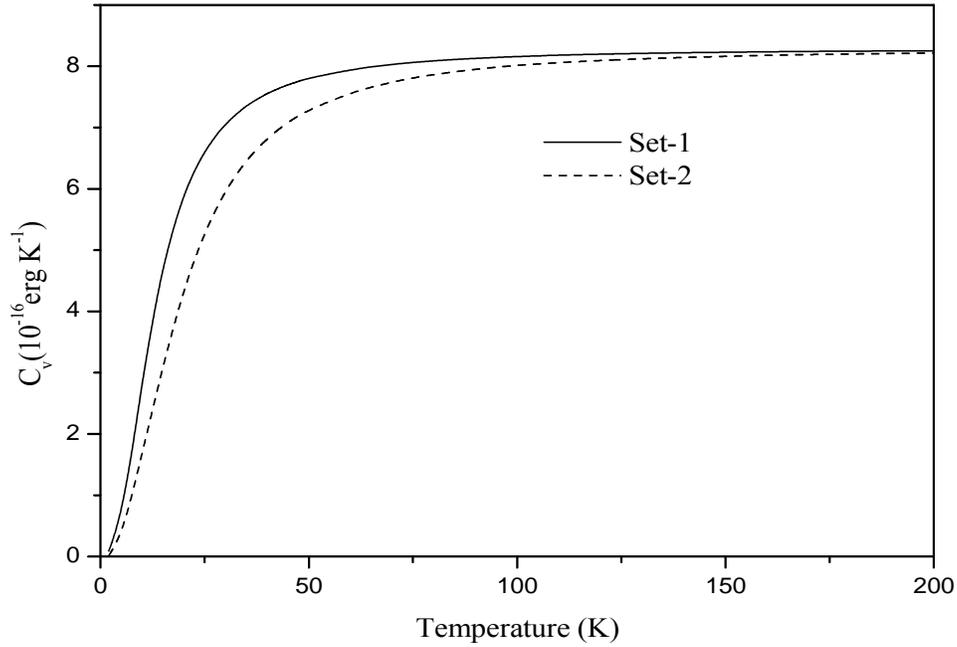

Fig. 5 Variation of Specific heat with temperature for $RbC_{60}$.

For the sake of discussion we have further calculated thermodynamic quantities for both the sets. The calculations with set-2 may be justified, as the range of phonon energies is not far from experimental values. Fig. 5 shows the variation of calculated specific heat with temperature. This calculation takes into account phonon frequencies at 48 points in the Brillouin zone of NaCl structure. Kellermann earlier stressed that these are enough to calculate specific heat, thermal expansion etc. Since the expression for phonon DOS cannot be found as a function of $\omega$. With large number of phonon frequencies one can approximate it numerically and Eq's. (20-23) are extensively used to derive the thermodynamic quantities.

From Fig.5 it is implied that for set-1 the maximum at lower energy is much more prominent compared to set-2 due to the lowering of optical phonons which falls in energy range of acoustic phonons. This results in more contribution to the specific heat at temperatures lower than the room temperature. At temperatures higher than 250 K both the calculations approach classical value 6 K. Fig's. 6 and 7 show the variation of volume thermal expansion and thermal expansion coefficient respectively.

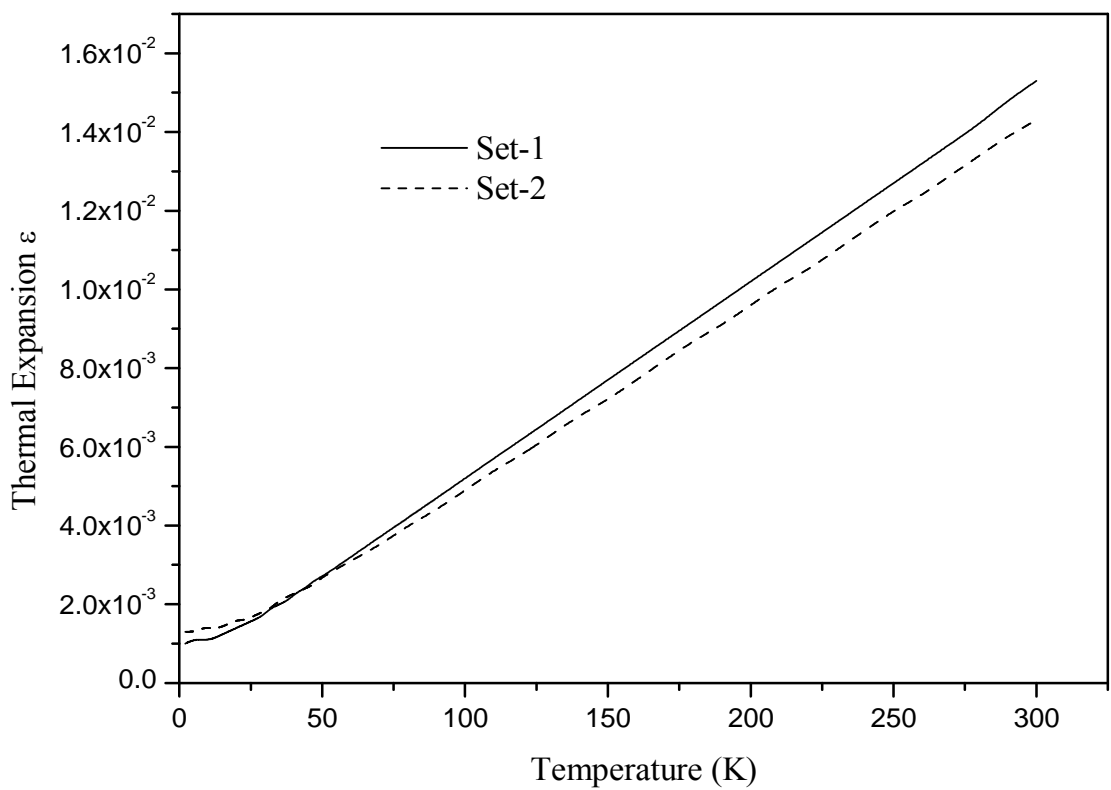

Fig. 6 Variation of volume thermal expansion ε of RbC$_{60}$ with temperature.

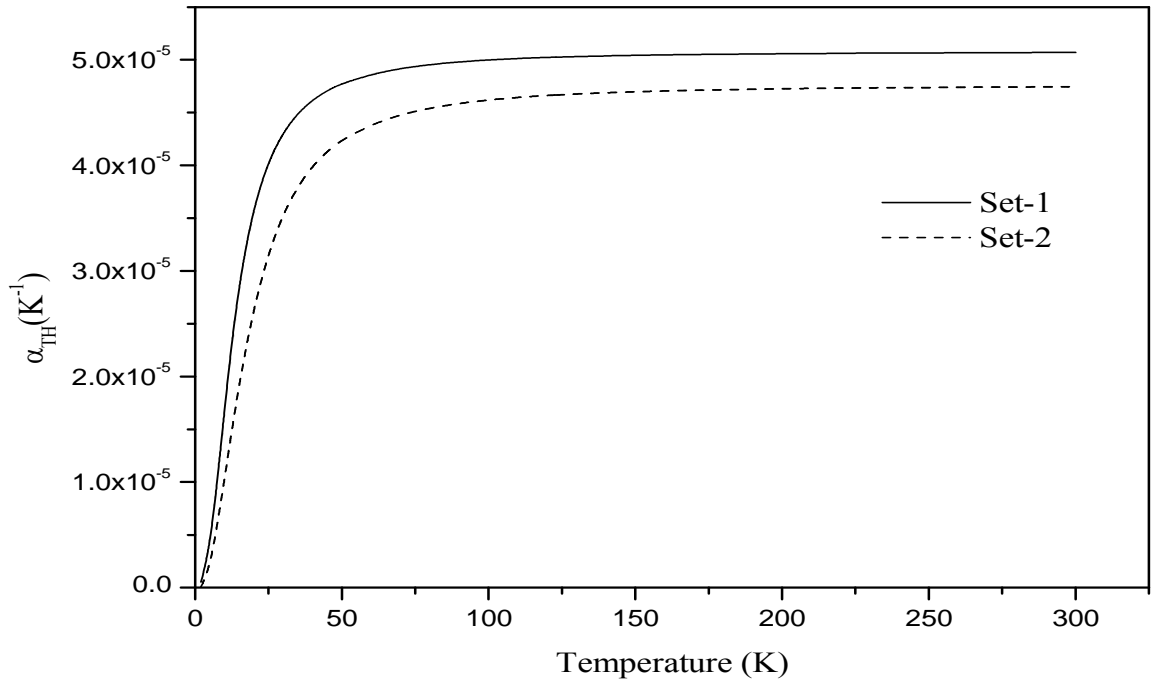

Fig.7. Variation of volume thermal expansion coefficient $\alpha_{TH}$ of RbC$_{60}$ with temperature

From Eq. 21, it is clear that thermal expansion is directly proportional to Gruneisen parameter and inversely proportional to bulk modulus. In case of RbC$_{60}$ for set-2, bulk modulus is smaller than for set-1 and Gruneisen parameter is also small. These two quantities compete with each other and result in cross over at about 50 K. In the low temperature region bulk modulus determines the thermal expansion as the DOS is small for both, which nullifies the contribution of Gruneisen parameter. Above 50 K the DOS for two sets undergoes a change and hence the Gruneisen parameter (larger for set-1) becomes important and hence thermal expansion increases for set-1. The calculation for volume thermal expansion coefficient shows the expected behavior. The larger value for set-1 is due to greater value of Gruneisen parameter. Similar calculations have been performed for Na, K and Cs doped C$_{60}$ solid in the same phase and same stoichiometry. The dispersion curves show similar characteristics. Fig's 8-10 show the phonon DOS curves for different alkali metals.

From the Fig's.8-10 it is clear that the cut-off frequency gradually decreases from about 12 T Hz for Na to 10 T Hz for Cs with set-1 calculations. Similar dependence is seen with set-2 calculations. Not only this, with increase in size of Alkali metal the gap in the phonon spectrum also decreases for set-2. It has significant effect on the calculated thermodynamic quantities. The average value of the Gruniesen parameter, elastic constants and bulk modulus are given in Table 6 and 7 for set-1 and set-2 respectively.

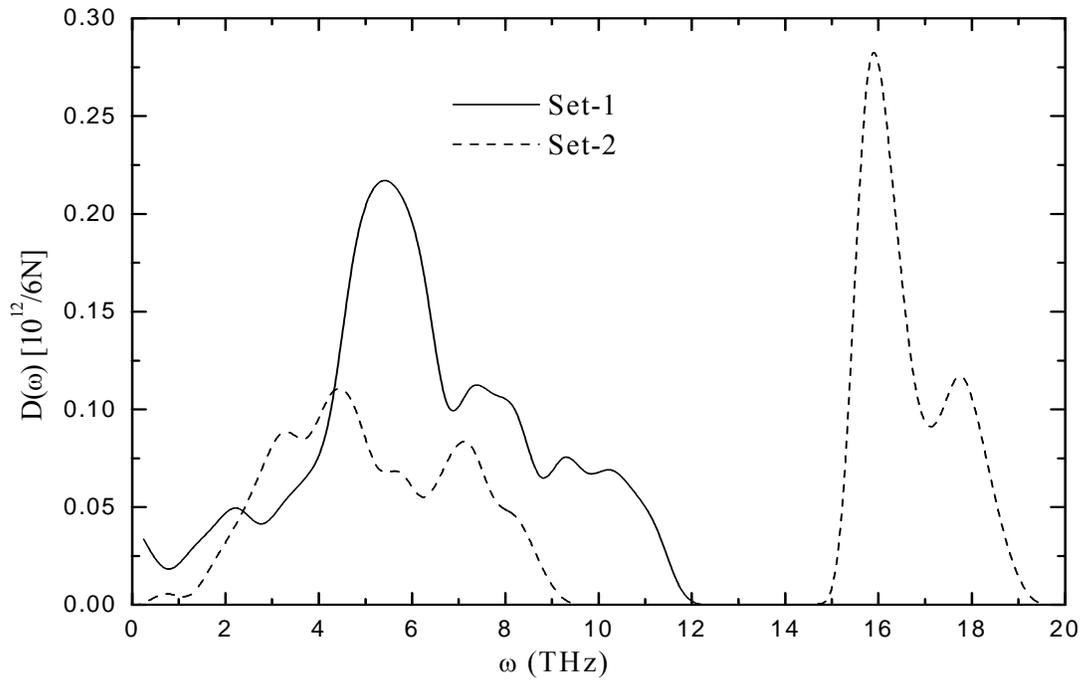

Fig. 8. Density of states curve for $NaC_{60}$ in the FCC phase.

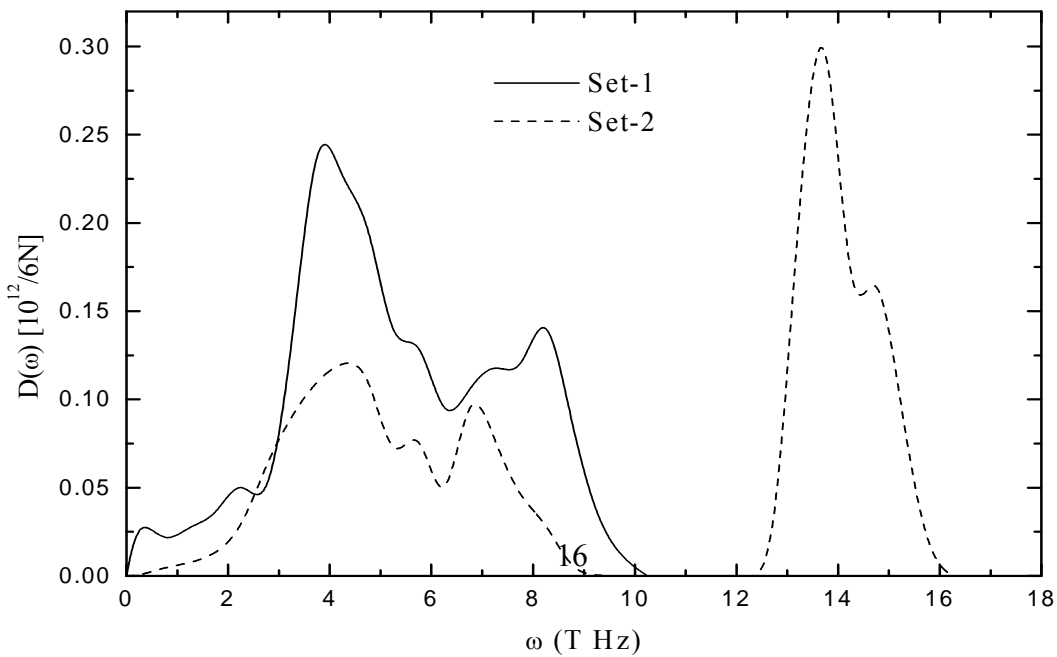

Fig.9. Density of states curve for $KC_{60}$ in the FCC phase

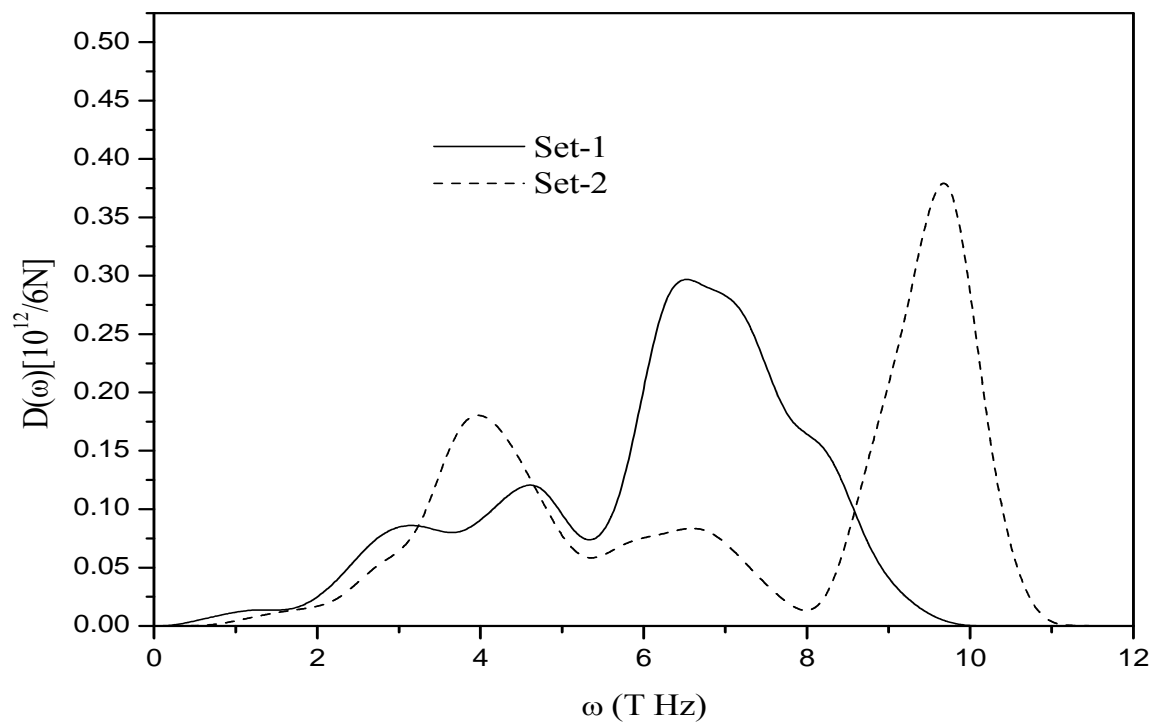

Fig. 10. Density of states curve for $CsC_{60}$ in FCC phase.

**Table No. 6 Thermodynamic quantities for set-1**

|  | Average Gruneisen parameter | Elastic Constants ($10^{10}$ dyne/cm$^2$) | | | Bulk Modulus ($10^{10}$ dyne/cm$^2$) |
|---|---|---|---|---|---|
|  |  | $C_{11}$ | $C_{12}$ | $C_{44}$ |  |
| $NaC_{60}$ set-1 | 6.46 | 27.60 | 6.91 | 7.12 | 13.81 |
| $KC_{60}$ set-1 | 6.96 | 26.67 | 6.90 | 7.14 | 13.49 |
| $RbC_{60}$ set-1 | 5.88 | 28.06 | 6.92 | 7.12 | 13.97 |
| $CsC_{60}$ set-1 | 5.02 | 28.92 | 6.03 | 6.00 | 13.66 |

**Table No. 7 Thermodynamic quantities for set-2**

|  | Average Gruneisen parameter | Elastic constants ($10^{10}$ dyne/cm$^2$) | | | Bulk Modulus ($10^{10}$ dyne/cm$^2$) |
|---|---|---|---|---|---|
|  |  | $C_{11}$ | $C_{12}$ | $C_{44}$ |  |
| NaC$_{60}$ set-2 | 5.10 | 26.84 | 5.23 | 5.32 | 12.43 |
| KC$_{60}$ set-2 | 5.01 | 26.49 | 5.01 | 4.67 | 12.17 |
| RbC$_{60}$ set-2 | 4.99 | 27.71 | 4.79 | 4.57 | 12.43 |
| CsC$_{60}$ set-2 | 5.00 | 28.31 | 4.21 | 4.08 | 12.24 |

From Table 6, we infer that average **Gruneisen** parameter decreases with increase in size of the atom and hence thermal expansion coefficient also decreases accordingly. The effect of phonon

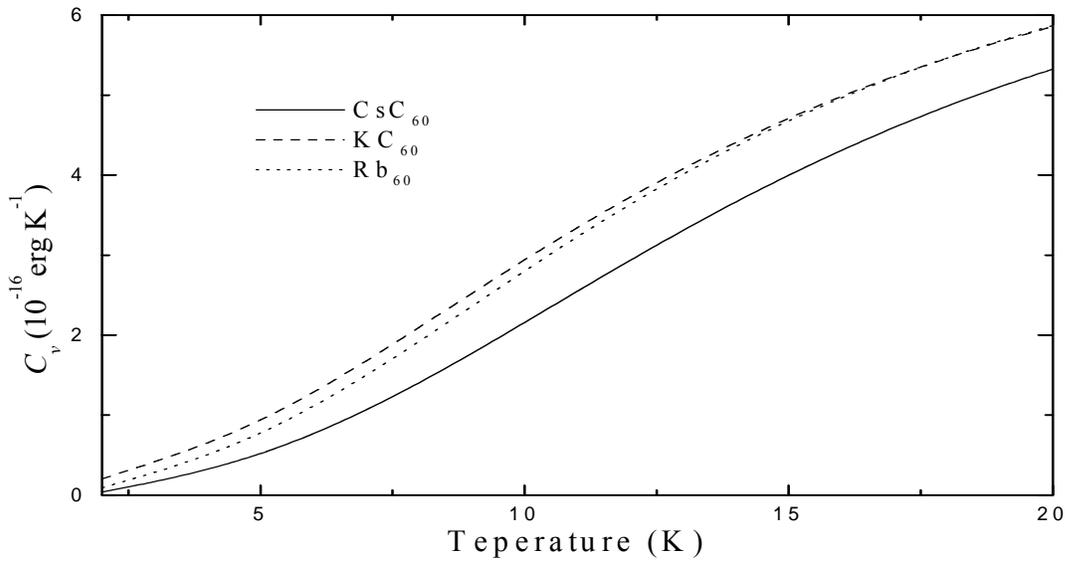

Fig.11 Variation of Specific heat with temperature for set-1.

DOS curve is manifested in Fig. 11. $C_v$ curve for CsC$_{60}$ is widely separated from other two (Rb and K), which is due to spreading of density of states in CsC$_{60}$ (Fig. 10) over a large range compared to RbC$_{60}$ and KC$_{60}$ (Fig. 1 and Fig. 9). The variation of $C_v$ and thermal expansion coefficient $\alpha_{TH}$ with temperature is illustrated in Fig. 11 and Fig 12. The value of thermal expansion coefficient for pure C$_{60}$ is $6.1\times10^{-5}$ K$^{-1}$[19]. From Fig. 12, it can be inferred that presently calculated values are of the same order. The volume thermal expansion coefficient deceases with increase in size of the alkali metal doped in C$_{60}$ solid. Therefore doping decreases anharmonicity in general, and further it decreases with increase in size of the alkali metal.

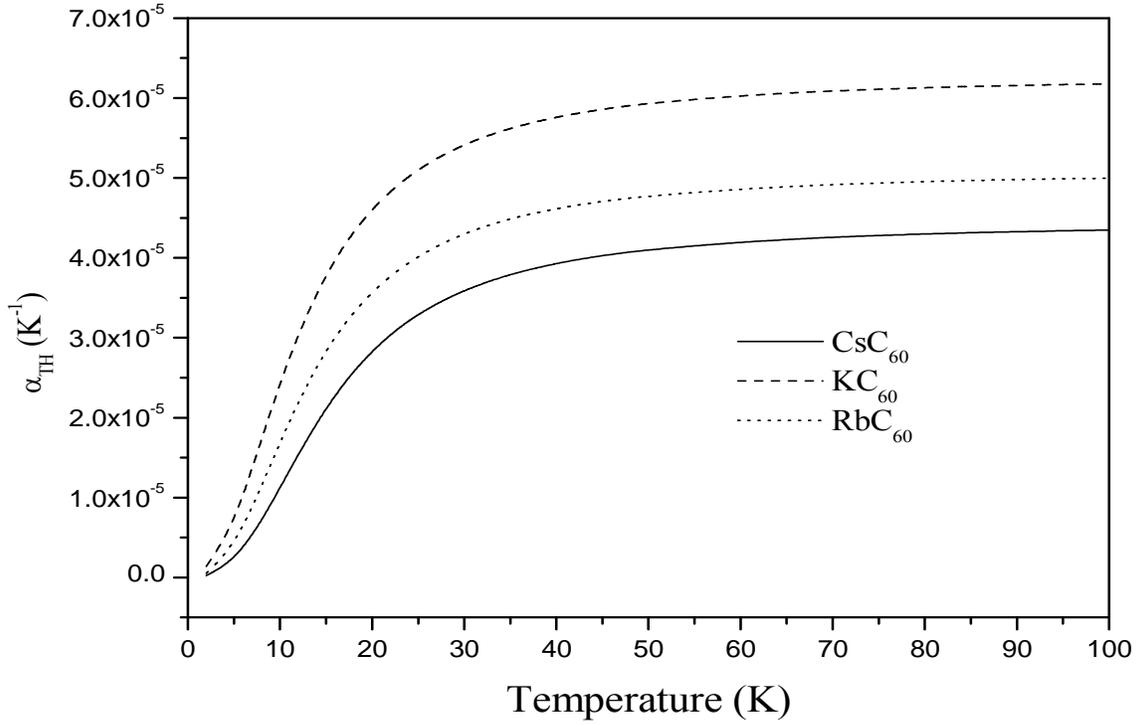

Fig. 12 Variation of thermal expansion coefficient for $MC_{60}$ solids in FCC phase with temperature.

## 4. Concluding remarks

We conclude that the form of the potential suggested in the earlier work accounts well for the lattice dynamical and thermodynamic properties of FCC $MC_{60}$ solids as well. We could not compare (due to the lack of experimental work reported in the literature) the calculation of thermodynamic quantities with experimental values except for DOS in $RbC_{60}$. But phonon DOS (set-1) in $RbC_{60}$ presented here resemble qualitatively with neutron scattering results. The consistency is attributed to the fact that various interactions between $C_{60}$ and alkali metal are properly incorporated within the framework of RIM. From DOS curves it is clear that alkali metal-carbon interaction is instrumental in deciding the form of these curves and hence the thermodynamic quantities. We expect qualitative agreement of calculated quantities with experimental values as the model of interactions is quite simple. A comparison has been made with another set of parameters (set-2) to discuss the effect of lattice expansion and change in the phonon density of states. The present work reported here for several thermodynamic quantities motivate more experimental work in doped $C_{60}$.

## Acknowledgement

RK would like to acknowledge the use of LAPACK subroutine (LAPACK driver routine (version 3.0) Univ. of Tennessee, Univ. of California Berkeley, NAG Ltd., Courant Institute, Argonne National Lab, and Rice University, June 30, 1999) to diagonalize the Dynamical matrix.

## References

1. Q. Zhu, O. Zhou, J. E. Fischer, A. R. McGhie, W. J. Romanov, R. M. Strongin, M. A Cichey, A. B. Smith III, Phys. Rev. B **47** (1993) 13948.
2. O. Chauvet, G. Oszlanyi, L. Forro, P. W. Stephens, M. Tegze, G. Faigel, A Janossy, Phys. Rev. Lett. **72** (1994) 721.
3. H. M. Guerrero, R.L. Cappelletti, D.A. Neumann, Tanner Yildirim, Chem. Phys. Lett. **297** (1998) 265.
4. Q. Zhu, D. E. Cox, J. E. Fischer, Phys rev. B **51** (1995) 3966.
5. L. Granasy, T. Kemeny, G. Oszlanyi, G. Bortel, g. Faigel, M. Tegze, S. Pekker, L. Forro, A. Jannosy, Solid State Comm. **97** (1996) 573.
6. K. Ranjan, K. Dharamvir and V. K. Jindal, Physica B **365** (2005) 121.
7. K. Ranjan, K. Dharamvir and V. K. Jindal, Indian J. of Pure & Applied Physics **43** (2005) 654.
8. K. Ranjan, Keya Dharamvir and V. K. Jindal, Physica B **371** (2006) 232
9. E. W. Kellermann, Phil. Trans. Roy. Soc. (London) **238** (1940) 513.
10. B. Renker, H. Schober, F. Gompf, R. heid and E. Ressouche, Phys. Rev. B **53** (1996) R14701
11. R. K. Singh, Physics Reports **85**(1982) 259
12. A. A. Maradudin, Theory of Lattice Dynamics in the Harmonic Approximation, IInd edition, Academic Press-1971.
13. A. D. B. Woods, W. Cochran and B. N. Brockhouse, Phys. Rev. B **119**(1960) 980
14. V. K. Jindal and J. Kalus, J. Phys. C 16, (1983) 3061; Phys. Stat. Sol. (b) **133** (1986) 189
15. Yang Wang, David Tomanek, George F. Bertsch and Rodney S. Ruoff, Phys. Rev. B 47 (1993) 6711

16. Q. Zhu, O. Zhou, J. E. Fischer, A. R. McGhie, W. J. Romanow, R. M. Strongin, M. A. Cichy, and A. B. Smith III, Phys. Rev. B **47** (1993) 13948

17. M. S. Dresselhaus, G. Dresselhaus and P. C. Eklund, Science of Fullerenes and Carbon Nanotubes (Acad. Press Inc., NY, 1996) page 457

18. B. Renker, F. Gompf, H. Schober, P. Adelmann, H. J. Bornemann, and R. Heid, Z. Phys. B **92** (1993) 451.

19. M. S. Dresselhaus, G. Dresselhaus and P. C. Eklund, Science of Fullerenes and Carbon Nanotubes (Acad. Press Inc., NY, 1996) page 172